\documentclass{ECOC-author}
\usepackage{amsmath,amssymb}
\usepackage{tikz,pgfplots}
\usepackage{subfig}
\usepackage{floatrow}
\usepackage{scalerel}
\usetikzlibrary{backgrounds,shapes,arrows,positioning,calc}
\usepackage[pdftex,colorlinks=true,bookmarks=false,citecolor=blue,urlcolor=blue]{hyperref} 
\pgfplotsset{compat=1.10}
\pgfdeclarelayer{background}
\pgfdeclarelayer{foreground}
\pgfsetlayers{background,main,foreground}
\newfloatcommand{capbtabbox}{table}[][0.9\FBwidth]

\begin{document}
\title{A NOVEL SOFT-AIDED BIT-MARKING DECODER FOR PRODUCT CODES}
\author{Gabriele Liga\ad{1}\corr, Alireza Sheikh\ad{1}, and Alex Alvarado\ad{1}}
\address{\add{1}{Department of Electrical Engineering, Eindhoven University of Technology, Eindhoven, The Netherlands} \email{g.liga@tue.nl}}

\begin{abstract}
We introduce a novel soft-aided hard-decision decoder for product codes adopting bit marking via updated reliabilities at each decoding iteration. Gains up to 0.8 dB vs.~standard iterative bounded distance decoding and up to 0.3 dB vs.~our previously proposed bit-marking decoder are demonstrated.  
\end{abstract}

\maketitle

\section{Introduction}
As optical line rates are quickly shifting towards 400 Gbit/s and beyond, implementing soft-decision forward-error correction (SD-FEC) decoders becomes increasingly challenging due to their high decoding complexity and internal data-flow. 
For this reason, lately, considerable attention has been devoted to hard-decision FEC (HD-FEC) schemes whose decoding complexity and data-flow can be order of magnitudes less than SD-FEC.

In the context of HD-FEC, product codes (PCs) \cite{Elias1954} (and generalisation thereof, such as staircase codes \cite{Smith2012}) offer at least at high coding rates a very good compromise between complexity and performance. This is due to the use of iterative algebraic decoding which is well known to be capacity-approaching at high rates in a binary-symmetric channel. A product code constructed on a $(n,k)$ component code is a set of square arrays of size $n\times n$ where any row and column of each array is a valid codeword in the component code. Typically, Bose-Chaudhuri-Hocquenghem (BCH) or Reed-Solomon codes are used as a component code. 

PCs based on BCH codes are typically decoded using iterative bounded distance decoding (iBDD) which can lead to substantial coding gains. However, the main performance limitation of BDD is due to so-called decoding \emph{failures} and \emph{miscorrections}. A decoding failure is declared when the received vector of bits is not within the correction capability of the component code. A miscorrection instead occurs when the received vector is succesfully decoded, but it is mapped into a codeword different from the transmitted one. 
 
A number of new decoding strategies have recently been proposed to tackle both failures and miscorrections, and thus, improve the performance of iBDD decoders whilst preserving an iterative algebraic decoding structure \cite{Hager2017, Sheikh2018ECOC, Sheikh2019, Lei2018ISTC, Sheikh2018ISTC, FougstedtOFC2018}. Except from \cite{Hager2017}, all these strategies somehow require the use of soft information from the channel. For instance in \cite{Sheikh2018ECOC,Sheikh2019}, the channel bit reliabilities are updated at each iteration based on the previous BDD output and are used to produce new hard decisions for the next BDD decoding stage.    

In our previous work \cite{Lei2018ISTC, Lei2019arXiv}, we showed how marking the input bits using the log-likelihood ratios (LLRs) can effectively prevent failures and miscorrections resulting in a significant improvement of the iBDD performance. We referred to this approach as \emph{soft-aided bit-marking} (SABM). SABM was first introduced in \cite{Lei2018ISTC} for staircase codes, and later also applied to PCs \cite{Lei2019arXiv}.   

In this contribution, we extend the SABM algorithm to include scaled reliabilities (SRs) in the decoding process. Differently from \cite{Sheikh2018ECOC, Sheikh2019}, the SRs are used to update the bit-marking process at each decoding iteration. The proposed design efficiently merges SABM and SRs to optimise the decoding of PCs. By doing that, we achieve up to 0.8 dB gain compared to standard iBDD and 0.3 dB compared to the previous implementation of SABM. 

\section{Soft-aided bit marking decoder}
\label{sec:SABM}
\begin{figure}[!t]
\centering
\includegraphics[scale=0.16]{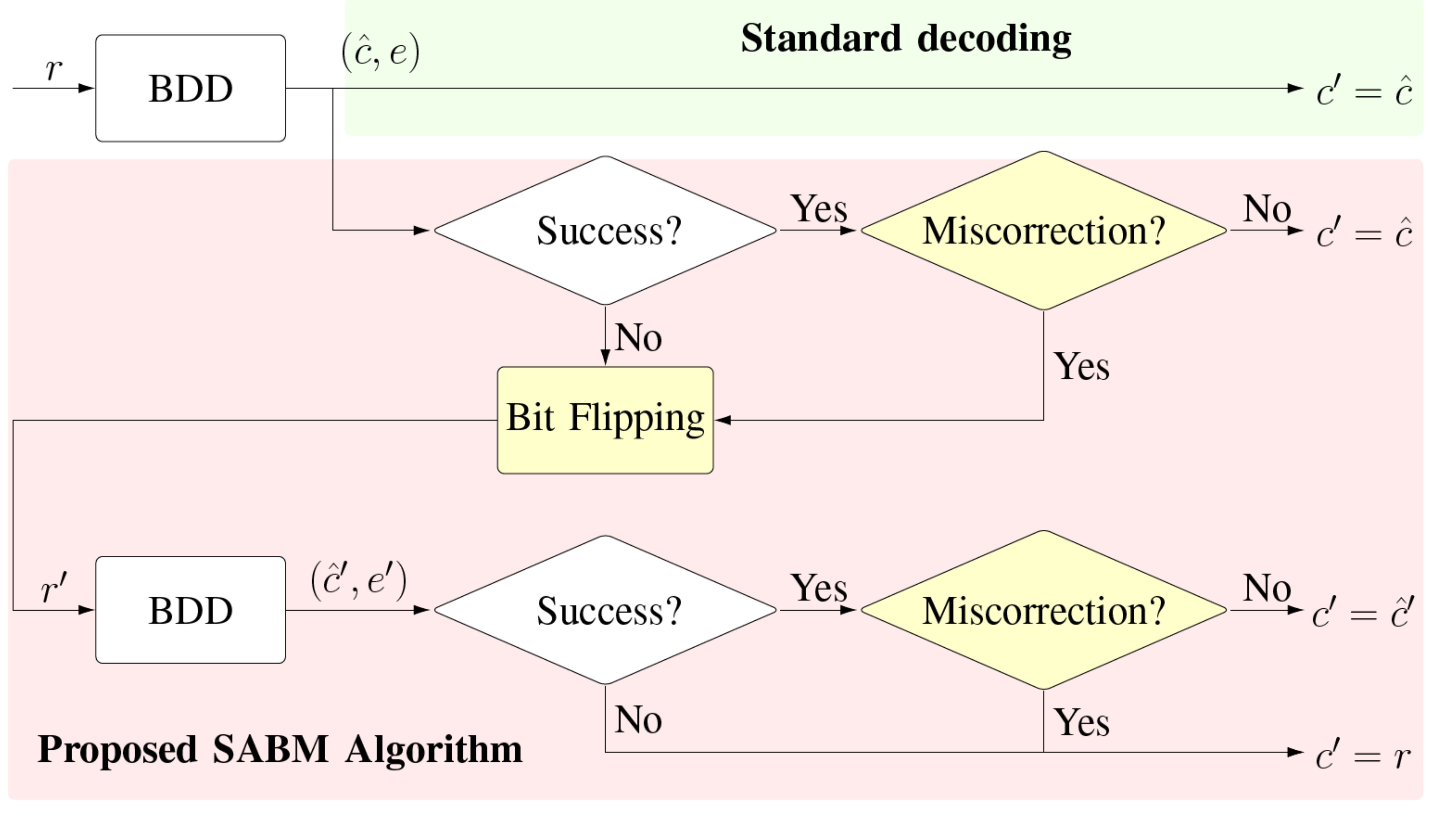}
\caption{Workflow of the SABM algorithm presented in \cite{Lei2018ISTC}.}
\label{fig:SABM}
\end{figure}

The SABM algorithm was shown to be an effective way to mitigate failures and miscorrections in iBDD \cite{Lei2019arXiv}. The algorithm workflow is illustrated in Fig.~\ref{fig:SABM}. In the standard BDD (top part of Fig.~\ref{fig:SABM}), the received codeword $r$ is succesfully decoded if there exists a codeword $\hat{c}$ that differs by at most $t$ positions from $r$, where $t$ is the correction power of the component code. Alternatively, a failure is declared and the decoder simply outputs $c^{\prime}=r$. A successful BDD decoding does not always guarantee that the transmitted codeword is correctly recovered. If more than $t$ errors are introduced by the channel, and $r$ is within $t$ positions from a codeword $c^{\prime}\neq c$, the decoder will select $c^{\prime}$ as opposed to $c$, resulting in a so called \emph{miscorrection}. Miscorrections often result in the decoder adding more errors than the ones introduced by the channel, thus leading to a deterioration of the iBDD performance. 

SABM tackles both failures and miscorrections via a twofold action: \emph{miscorrection detection} and \emph{bit flipping}. This approach is enabled by the use of \emph{soft quantities} or \emph{reliabilities} extracted from the channel. In SABM, LLRs are used as a bit reliability measure. By setting a threshold in the LLR magnitude, bits are marked as highly reliable bits (HRBs) when they fall above such a threshold. Alternatively, they are labelled as highly unreliable bits. Two rules are adopted to prevent a miscorrection: i) the BDD decoder can never flip a bit from a previously succesfully decoded codeword; ii) HRBs can never be flipped. When either a failure occurs or a miscorrection is detected (red area in Fig.~\ref{fig:SABM}) a new decoding attempt is performed after \emph{bit flipping}. 

Bit flipping is used to facilitate the success of an additional decoding attempt. The LLRs are ordered based on their magnitude and the least reliable bit is flipped in the case of failure. In the case of a detected miscorrection, the least $d_{min}-t-w_H(e)$ bits are flipped, where $d_{min}$ and $w_H(e)$ are the minimum Hamming distance of the component code and the Hamming weight of the error vector $e$, respectively. If after an additional decoding attempt on the flipped received vector $r^{\prime}$ a further failure or miscorrection is detected, the output is set to $c^{\prime}=r$ and passed to the next BDD decoding stage. This process is performed for a few iterations. For PCs the number of these iterations is lower than the number of total iBDD iterations and can be optimised in terms of post-FEC bit-error rate (BER) performance (see Sec.~\ref{sec:Results}).

\section{SABM with scaled reliabilities}
\label{sec:SABM-SR}
In SABM the bit marking occurs only once (based on the channel LLRs) and is then used over a certain number of iterations. This approach is, however, sub-optimal as the marking mask loses its validity after bits are updated at every iteration by the BDD decoders.   
This problem can be solved by re-marking the output bits at each BDD iteration by defining a reliability measure for the outbound BDD decisions. The SR measure proposed in \cite{Sheikh2018ECOC,Sheikh2019} linearly combines the BDD decisions with the channels LLRs. We call this new approach SABM-SR.

\begin{figure}[!t]
\scalebox{0.9}{\resizebox{0.95\columnwidth}{!} 
{\begin{tikzpicture} 
\tikzset{fontscale/.style = {font=\tiny}}
\tikzstyle{block} = [draw, rectangle, minimum height=0.8cm, minimum width=2.3cm, rounded corners=0.1cm, align=center,font=\tiny] 
\tikzstyle{Cir} = [draw, circle, minimum size=0.25em,font=\tiny] \node[block,fill=red!20] (BDD1) at (-30pt,0) {SABM \\ row decoder}; 
\node[block,fill=blue!20,right=0.7cm of BDD1,] (BDD2) {SABM \\ column decoder}; 
\node[block, minimum height=2em, minimum width=2em, align=center, fill=yellow!20] (Mark1) at (40pt,45pt) {Bit \\ Marking}; 
\node[block, minimum height=1.5em, minimum width=2.5em, align=center, fill=yellow!20] (Mark2) at (-47pt,45pt) {Bit \\ Marking}; 
\node[block, minimum height=2em, minimum width=3.6em,fill=green!20] (SR1) at (-10pt,45pt) {}; 
\node[block, minimum height=2em, minimum width=3.6em,fill=green!20] (SR2) at (75pt,45pt) {}; 
\node[Cir,inner sep=0pt,fill=white] (mult) at (-18pt,45pt) {$\times$};
\node[Cir, right=0.3 cm of mult, inner sep=0pt,fill=white] (add) {$+$}; 
\draw[->] (BDD1.north-|mult)--++(0pt,10pt)node[right]{$\scaleto{u^{\text{r}}_{i,j}}{8pt}$}--(mult); 
\draw[->] (Mark2)--++(0pt,-24pt)node[right]{$\scaleto{\psi^{\text{r}}_{i,j}}{8pt}$}--(Mark2|-BDD1.north); 
\draw[->] (Mark1)--++(0pt,-24pt)node[right]{$\scaleto{\psi^{\text{c}}_{i,j}}{8pt}$}--(Mark1|-BDD2.north); 
\draw[->] (Mark2)--(Mark2|-BDD1.north); 
\node[above=0.35cm of add] (llr1) {$\scaleto{l_{i,j}}{8pt}$}; 
\node[above=0.35cm of mult] (w) {$\scaleto{w^{\text{r}}}{5pt}$}; 
\draw[->] (llr1)--(add); 
\draw[->] (w)--(mult); 
\draw[->] (mult)--(add); 
\draw[->] (BDD1.east)--++(10pt,0pt)node[above]{$\scaleto{\hat{b}_{i,j}}{8pt}$}--(BDD2); 
\draw[->] (add)--++(18pt,0pt)node[above]{$\scaleto{\phi_{i,j}}{8pt}$}--(Mark1); 
\node[Cir,inner sep=0pt,fill=white] (mult2) at (67pt,45pt) {$\times$}; 
\node[Cir, right=0.3 cm of mult2, inner sep=0pt,fill=white] (add2) {$+$}; 
\draw[->] (BDD2.north-|mult2)--++(0pt,10pt)node[right]{$\scaleto{u^{\text{c}}_{i,j}}{8pt}$} --(mult2); 
\node[above=0.35cm of add2] (llr2) {$\scaleto{l_{i,j}}{8pt}$}; 
\node[above=0.35cm of mult2] (w2) {$\scaleto{w^{\text{c}}}{5pt}$}; 
\draw[->] (llr2)--(add2); 
\draw[->] (w2)--(mult2); 
\draw[->] (Mark2)++(-25pt,0pt)--(Mark2); 
\draw[->] (BDD1)++(-40pt,0pt)--(BDD1); 
\draw[->] (mult2)--(add2); 
\draw[->] (add2)--++(15pt,0pt); 
\draw[->] (BDD2)--++(42pt,0pt); 
\node[] (end) at (80pt,0pt) {};
\end{tikzpicture} }}
\caption{Schematic diagram of one iteration of the proposed SABM-SR algorithm. }
\label{fig:SABM-SR}
\end{figure}
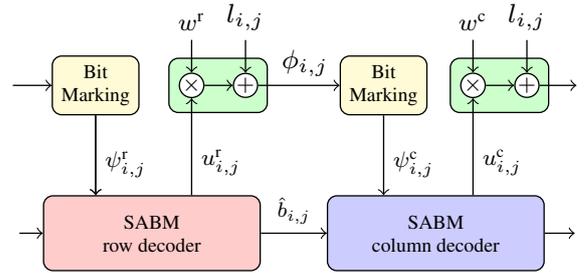

The SABM-SR algorithm is illustrated in Fig.~\ref{fig:SABM-SR}. The BDD output can provide at each iteration quantised information by assigning values -1 and +1 to successfully decoded bits 0 and 1, respectively. When a failure occurs the BDD output will instead be identically set to 0. The green blocks in Fig.~\ref{fig:SABM-SR} show how the reliabilities are updated at each iteration for both row and column decoding. Let $u^{\text{r}}_{i,j} \in \{-1,0,+1\}$ be the quantised reliability information provided by the row BDD decoder for the bit in row $i$ and column $j$. The SR after row decoding at each iteration is defined as \cite{Sheikh2019}

\begin{equation}
\phi_{i,j}=w^{\text{r}}u^{\text{r}}_{i,j}+l_{i,j}  \quad \forall \; i,j=1,2,...n
\label{eq:SR}
\end{equation}
where $w^{\text{r}} \in \mathbb{R}^+$ are optimised weights and $l_{i,j}$ are the LLRs. 

As illustrated in Fig.~\ref{fig:SABM-SR} (yellow blocks), the SRs $\phi_{i,j}$ can be used to update the SABM marking stage. The bit-marking block takes as an input the SRs from the previous (column) decoding iteration and yields a new set of marked bits $\psi^{\text{r}}_{i,j}$. The new marked bits are then passed to the SABM row decoder which produces a block of output bits $\hat{b}_{i,j}$ and a corresponding block $u^{\text{r}}_{i,j}$. This information is used to update again the reliabilities $\phi_{i,j}$ according to \eqref{eq:SR}. A marking update is then performed via the $\phi_{i,j}$ for the SABM column decoder.         
Differently from \cite{Sheikh2018ECOC, Sheikh2019}, here the reliability information at each iteration is used merely to re-mark the bits at the input of the next BDD decoder, and the BDD output $\hat{b}_{i,j}$ is still used as the input of the next BDD decoder. 



\section{Results and Discussion}
\label{sec:Results}
The two decoders described in this paper were numerically assessed in an additive white Gaussian noise channel with binary antipodal modulation (2PAM) and noise power spectral density $N_0/2$. Ten decoding iterations were used for all the implemented decoders. For SABM, and SABM-SR, the reliability threshold and the number of iterations where miscorrection detection (and bit flipping) is performed was optimised and set to 5 and 5 (out of 10) iterations, respectively. The SR weights were numerically optimised 
and it was found $w^{\text{r}}\approx w^{\text{c}}$. The optimised vector $\boldsymbol{w}\triangleq [w_1,w_2,...,w_N]$, where $w_k$ is the SR weight at iteration $k$ and $N$ is the number of marking iterations, was found to be $\boldsymbol{w}=[3.42, 3.87, 4.08, 4.27, 4.49]$.

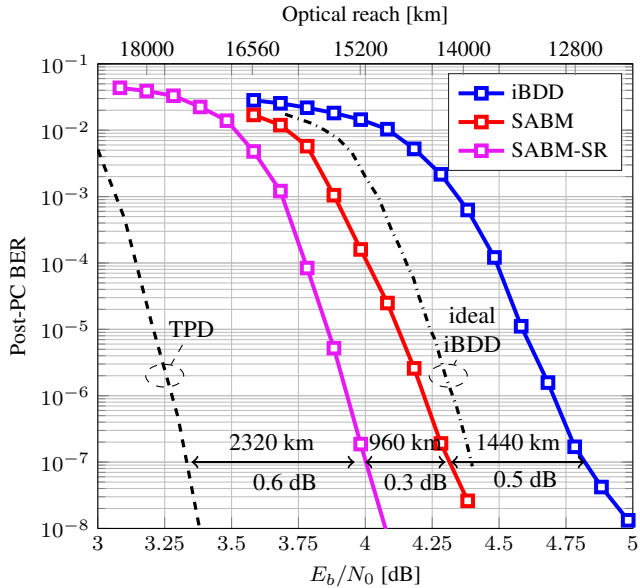
\begin{figure}[!t]
    \centering
%
%
\definecolor{mycolor1}{rgb}{0.81472,0.63236,0.95751}%
\definecolor{mycolor2}{rgb}{0.90579,0.09754,0.96489}%
\definecolor{mycolor3}{rgb}{0.12699,0.27850,0.15761}%
\definecolor{mycolor4}{rgb}{0.91338,0.54688,0.97059}%
\begin{tikzpicture}

\begin{axis}[%
width=0.8\columnwidth,
height=0.7\columnwidth,
at={(0.766in,0.486in)},
scale only axis,
xmin=3,
xmax=5,
ymode=log,
ymin=1e-8,
ymax=0.1,
yminorticks,
axis background/.style={fill=white},
xmajorgrids,
ymajorgrids,
yminorgrids,
xtick={3,3.25,...,5},
ylabel={Post-PC BER},
xlabel={$E_b/N_0$ [dB]},
legend style={legend cell align=left, align=left, draw=white!15!black}
]
\addplot [color=blue, line width=1.5pt, mark=square*, mark options={solid,blue,fill=white}]
  table[row sep=crcr]{%
3.58263052328897	0.0281964131881901\\
3.68263052328897	0.025451484062965\\
3.78263052328897	0.0218388284125617\\
3.88263052328897	0.0181744850810557\\
3.98263052328897	0.0144498394549299\\
4.08263052328897	0.010375127261336\\
4.18263052328897	0.00525021536533793\\
4.28263052328897	0.00216500900618686\\
4.38263052328897	0.000628866786749158\\
4.48263052328897	0.000121074477249589\\
4.58263052328897	1.11142460818528e-05\\
4.68263052328897	1.57710367372668e-06\\
4.78263052328897	1.70058664008983e-07\\
4.88263052328897	4.20914459090663e-08\\
4.98263052328897	1.32629192963929e-08\\
};
\addlegendentry{iBDD}


\addplot [color=red, line width=1.5pt, mark=square*, mark options={solid, red,fill=white}]
  table[row sep=crcr]{%
3.58263052328897	0.0170177774297126\\
3.68263052328897	0.0119375048946668\\
3.78263052328897	0.00574986294933041\\
3.88263052328897	0.00104967760461534\\
3.98263052328897	0.000159892447855483\\
4.08263052328897	2.49017945409001e-05\\
4.18263052328897	2.58268770852694e-06\\
4.28263052328897	1.91002233750485e-07\\
4.38263052328897	2.60702828756043e-08\\
};
\addlegendentry{SABM}

\addplot [color=mycolor2, line width=1.5pt, mark=square*, mark options={solid, mycolor2, fill=white}]
  table[row sep=crcr]{%
3.08263052328897	0.0435844623698019\\
3.18263052328897	0.0389537160310126\\
3.28263052328897	0.0330722844388754\\
3.38263052328897	0.0224191401049417\\
3.48263052328897	0.0139360952306367\\
3.58263052328897	0.0047877672488057\\
3.68263052328897	0.00121058814315921\\
3.78263052328897	8.37610975444792e-05\\
3.88263052328897	5.20241441221941e-06\\
3.98263052328897 1.86711340176776e-07\\
4.098263052328897 4.86711340176776e-09\\
};
\addlegendentry{SABM-SR}


\addplot [color=black, dashed, line width=1.2pt, forget plot]
  table[row sep=crcr]{%
2.9	0.0202655407357798\\
3	0.00509089398311894\\
3.1	0.000466080054785853\\
3.2	1.29053451418654e-05\\
3.3	4.71449588895136e-07\\
3.4	3.91569425992638e-09\\
};

\addplot [color=black, dashdotted, line width=1.2pt]
  table[row sep=crcr]{%
3.7	0.0174572097353834\\
3.8	0.0126907656900912\\
3.85	0.0099401617923037\\
3.9	0.00733737247168213\\
3.95	0.00458914850100261\\
4	0.001991\\
4.05	0.00095038654\\
4.1	0.000301315910354874\\
4.12	0.000206933197175074\\
4.14	0.00013657876\\
4.16	8.57511777036246e-05\\
4.18	5.90132862345964e-05\\
4.2	3.13869824607447e-05\\
4.22	1.86504142035067e-05\\
4.24	1.04626304372652e-05\\
4.26	7.065408069237e-06\\
4.3	1.952353e-06\\
4.34	7.09516704802563e-07\\
4.38	1.85795893e-07\\
4.4	8.6144412e-08\\
};

\draw[thick,<->] (axis cs:4.32,1e-7)--(axis cs:4.82, 1e-7);
\draw[thick,<->] (axis cs:4,1e-7)--(axis cs:4.3, 1e-7);
\draw[thick,<->] (axis cs:3.35,1e-7)--(axis cs:3.96, 1e-7);
\node[font=\small] at (axis cs:3.7,5e-8) {0.6 dB};
\node[font=\small] at (axis cs:3.65,2e-7) {2320 km};
\node[font=\small] at (axis cs:4.19,5e-8) {0.3 dB};
\begin{pgfonlayer}{foreground}
\node[font=\small] at (axis cs:4.15,2e-7) {960 km};
\end{pgfonlayer}
\node[font=\small] at (axis cs:4.6,5.5e-8) {0.5 dB};
\node[font=\small] at (axis cs:4.57,2e-7) {1440 km};
\node[font=\small] (tpd) at (axis cs:3.35,1e-5) {TPD};
\node[font=\small,align=center] (ibdd) at (axis cs:4.4,1e-5) {ideal \\ iBDD};

\node[font=\small,ellipse,dashed,draw,minimum height=0.3cm,minimum width=0.5cm] (ell1) at (axis cs:3.25,2e-6) {};
\node[font=\small,ellipse,dashed,draw,minimum height=0.3cm,minimum width=0.5cm] (ell2) at (axis cs:4.31,2e-6) {};
\draw[dashed] (tpd)--(ell1);
\draw[dashed] (ibdd)--(ell2);

\end{axis}
\begin{axis}
[axis x line* = top,
xmin=3,
xmax=5,
ymode=log,
ymin=1e-8,
ymax=0.1,
xtick={4.78454859,
4.36634053,
3.98089928,
3.57745292,
3.18300916
},
xticklabels={12800,14000,15200,16560,18000},
axis y line = none,
width=0.8\columnwidth,
height=0.7\columnwidth,
at={(0.766in,0.486in)},
scale only axis,
xlabel={Optical reach [km]},
xlabel near ticks
]%
\addplot []
  table[row sep=crcr]{%
3.7	1\\
3.8	1\\
3.85 1\\
3.9	1\\
3.95 1\\
4  1\\
4.05  1\\
};

\end{axis}

\end{tikzpicture}%
    \caption{BER vs. $E_b/N_0$ for the decoding schemes analysed and for a code rate $R$=0.78.}
    \label{fig:Results1}
\end{figure}

Fig.~\ref{fig:Results1} shows the post-PC bit-error rate (BER) as a function of $E_b/N_0$, where $E_b$ is the energy per information bit. The component code is in this case a BCH (7,2,1)\footnote{We use the common notation ($\nu$,$t$,$e$) for BCH codes, where $\nu$ is the order of the extended Galois field and $e$ number of bits of extension.}, yielding a PC rate $R=0.78$. Other than the decoding algorithms described in Secs.~\ref{sec:SABM}, \ref{sec:SABM-SR}, the performance of an ideal (miscorrection-free) iBDD (dash-dotted curve) and a \emph{fully-fledged} SD decoder such as the turbo product decoder (TPD) based on the Chase-Pyndiah algorithm \cite{Pyndiah1998}, are also shown as a reference.  

\begin{table}[!b]
\caption{Optical system paramaters.}
\begin{tabular}{cc}
\hline \hline
Parameter Name & Value \\
\hline \hline
Modulation Format & PM-QPSK \\
Symbol rate  & 33 GBaud \\
Channel Spacing & 33 GHz \\
No. of Channels & 295 \\
Roll-off factor & 0 \\
Fibre Type & Standard single-mode fibre \\
Span Length & 80 km \\
EDFA noise figure & 4.5 dB \\
\hline\hline
\end{tabular}
\label{table:parameters}
\end{table}

The SABM decoder (red curve) shows a 0.5 dB coding gain compared to iBDD (blue line) at a BER of $10^{-7}$. SABM also outperforms ideal iBDD (dashed-dotted black line), due to its capability to also mitigate failures. SABM-SR gains 0.3 dB over SABM and an overall 0.8 dB compared to iBDD. This substantial coding gain comes at expense of some additional decoding complexity, due to the SR calculation and the repeated marking operation. However, as noted in \cite{Sheikh2019} and also illustrated in Fig.~\ref{fig:SABM-SR}, the SRs are not passed between the SABM decoding stages and can be calculated and stored in place, considerably reducing the decoder data flow. The only messages passed to the decoders are a mask of marked bits and the previous decoder output (hard bits). Finally, TPD (dashed black line) gains 0.6 dB over SABM-SR. Thus, SABM-SR performance is closer to an SD decoder than to conventional HD decoding for PCs (iBDD).

Fig.~\ref{fig:Results1} also shows the performance in terms of achievable optical reach ($x$ axis at the top). In order to link $E_b/N_0$ to a given transmission distance, a closed-form for the enhanced Gaussian noise model proposed in \cite{Poggiolini2015} was used. For our analysis, we assumed a long-haul optical transmission system (parameters in Table \ref{table:parameters}) using polarisation-multiplexed quadrature phase-shift keying (PM-QPSK) and Erbium-doped fibre amplifiers (EDFAs). SABM yields a 1440 km reach increase, whilst SABM-SR gains an additional 960 km and an overall 2400 km compared to iBDD. Using an SD decoder (TPD) increases the reach by 2320 km compared to SABM-SR. 

\begin{figure}[!t]
    \centering
%
%
\definecolor{mycolor1}{rgb}{0.81472,0.63236,0.95751}%
\definecolor{mycolor2}{rgb}{0.90579,0.09754,0.96489}%
\definecolor{mycolor3}{rgb}{0.12699,0.27850,0.15761}%
\definecolor{mycolor4}{rgb}{0.91338,0.54688,0.97059}%
\begin{tikzpicture}

\begin{axis}[%
width=0.8\columnwidth,
height=0.7\columnwidth,
at={(0.766in,0.486in)},
scale only axis,
xmin=3.8,
xmax=5.2,
ymode=log,
ymin=1e-8,
ymax=0.1,
yminorticks=true,
axis background/.style={fill=white},
xmajorgrids,
ymajorgrids,
yminorgrids,
ylabel={Post-PC BER},
xlabel={$E_b/N_0$ [dB]},
legend style={legend cell align=left, align=left, draw=white!15!black},
]
\addplot [color=blue, line width=1.5pt, mark=square*, mark options={solid,blue,fill=white}]
  table[row sep=crcr]{%
4.09684128727424	0.0186322018171951\\
4.19684128727424	0.0167351762048984\\
4.29684128727424	0.0150685387160589\\
4.39684128727424	0.0132208819873602\\
4.49684128727424	0.0111606939654418\\
4.59684128727424	0.00895099875702456\\
4.69684128727424	0.00647152535844961\\
4.79684128727424	0.00354545613697239\\
4.89684128727424	0.000723201624621418\\
4.99684128727424	3.58250667243857e-05\\
5.09684128727424	6.16749668215983e-07\\
5.19684128727424	4.98015458001003e-09\\
5.29684128727424	0\\
5.39684128727424	0\\
5.49684128727424	0\\
};
\addlegendentry{iBDD}

\addplot [color=red, line width=1.5pt, mark=square*, mark options={solid, red,fill=white}]
  table[row sep=crcr]{%
4.09684128727424	0.0173662926069221\\
4.19684128727424	0.0138892876525271\\
4.29684128727424	0.0102422926769489\\
4.39684128727424	0.00538208364699498\\
4.49684128727424	0.00100024509374836\\
4.59684128727424	8.31568074788607e-05\\
4.69684128727424	7.81999402956069e-07\\
4.79684128727424	3.85166577178319e-09\\
4.89684128727424	0\\
4.99684128727424	0\\
5.09684128727424	0\\
5.19684128727424	0\\
5.29684128727424	0\\
5.39684128727424	0\\
};
\addlegendentry{SABM}

\addplot [color=mycolor2, line width=1.5pt, mark=square*, mark options={solid, mycolor2, fill=white}]
  table[row sep=crcr]{
3.39684128727424	0.0351557220636894\\
3.49684128727424	0.032829957458728\\
3.59684128727424	0.0312758880271704\\
3.69684128727424	0.0289893384219464\\
3.79684128727424	0.0263360235289998\\
3.89684128727424	0.0237651651756797\\
3.99684128727424	0.0194737487088811\\
4.09684128727424	0.0153251868839831\\
4.19684128727424	0.00842142119360655\\
4.29684128727424	0.00157385199838938\\
4.39684128727424	5.06474029346528e-05\\
4.49684128727424	1.78271286267877e-07\\
4.59684128727424	6.78271286267877e-10\\
};
\addlegendentry{SABM-SR}


\addplot [color=black, dashdotted, line width=1.2pt]
table[row sep=crcr]{%
4.87171729372303	2.99461745482507e-09\\
4.84171729372303	3.99461745482507e-08\\
4.81551091280726	2.23375624759708e-07\\
4.78963866089023	7.92925797770088e-07\\
4.76408976779176	3.42091503267974e-06\\
4.73885399279531	1.14074381253265e-05\\
4.71392158992996	3.47716330238721e-05\\
4.68928327607817	9.44600309377187e-05\\
4.66493020163605	0.000211770044872607\\
4.64085392348344	0.000481936949457965\\
4.61704638004776	0.000834419142455281\\
4.59349986826886	0.00148763079261222\\
4.5702070222925	    0.00221417220826454\\
4.54716079373821	0.00312344627821357\\
4.52435443340318	0.00410833942267401\\
4.50178147427775	0.004959932245799\\
4.47943571576071	0.00579574503053729\\
4.4573112089734	0.00655936028846232\\
};

\addplot[color=black, dashed, line width=1.2pt]
table[row sep=crcr]{%
		3.2	0.054437587082942\\
		3.3	0.046984287279358\\
		3.4	0.042677157346018\\
		3.5	0.035219971639152\\
		3.6	0.030124939456009\\
		3.7	0.0213965988332\\
		3.8	0.00829490177189\\
		3.9	0.000259890294429629\\
		3.925 7.4173e-05\\
		3.955 1.3443e-05\\
		3.9745 2.4169e-06\\
		4     1.6083e-07\\	
		4.02     1e-08\\	
	};

\draw[thick,<->] (axis cs:4.735,1e-7)--(axis cs:5.125, 1e-7);
\draw[thick,<->] (axis cs:4.515,1e-7)--(axis cs:4.73, 1e-7);
\draw[thick,<->] (axis cs:4.01,1e-7)--(axis cs:4.5, 1e-7);
\node[font=\small] at (axis cs:4.24,5e-8) {0.45 dB};
\node[font=\small] at (axis cs:4.24,2e-7) {1340 km};
\node[font=\small] at (axis cs:4.64,5e-8) {0.23 dB};
\node[font=\small] at (axis cs:4.62,2e-7) {640 km};
\node[font=\small] at (axis cs:5,5e-8) {0.4 dB};
\node[font=\small] at (axis cs:4.97,2e-7) {960 km};

\node[font=\small] (tpd) at (axis cs:4.07,1e-5) {TPD};
\node[font=\small,align=center] (ibdd) at (axis cs:4.89,1e-5) {ideal \\ iBDD};
\node[font=\small,ellipse,dashed,draw,minimum height=0.3cm,minimum width=0.5cm] (ell1) at (axis cs:3.97,2e-6) {};
\node[font=\small,ellipse,dashed,draw,minimum height=0.3cm,minimum width=0.5cm] (ell2) at (axis cs:4.79,2e-6) {};
\draw[dashed] (tpd)--(ell1);
\draw[dashed] (ibdd)--(ell2);
\end{axis}

\begin{axis}
[axis x line* = top,
width=0.8\columnwidth,
height=0.7\columnwidth,
at={(0.766in,0.486in)},
scale only axis,
xmin=3.8,
xmax=5.2,
ymode=log,
ymin=1e-8,
ymax=0.1,
xtick={5.09851735,4.79978352,4.48814060,    4.19524851,3.89209404},
xticklabels={10800,11520,12320,12800,13120,14000},
axis y line = none,
xlabel={Optical reach [km]},
xlabel near ticks
]%
\addplot []
  table[row sep=crcr]{%
3.7	1\\
3.8	1\\
3.85 1\\
3.9	1\\
3.95 1\\
4  1\\
4.05  1\\
};

\end{axis}

\end{tikzpicture}%
    \caption{BER vs. $E_b/N_0$ for the decoding schemes analysed and for a code rate $R$=0.87.}
    \label{fig:Results2}
\end{figure}
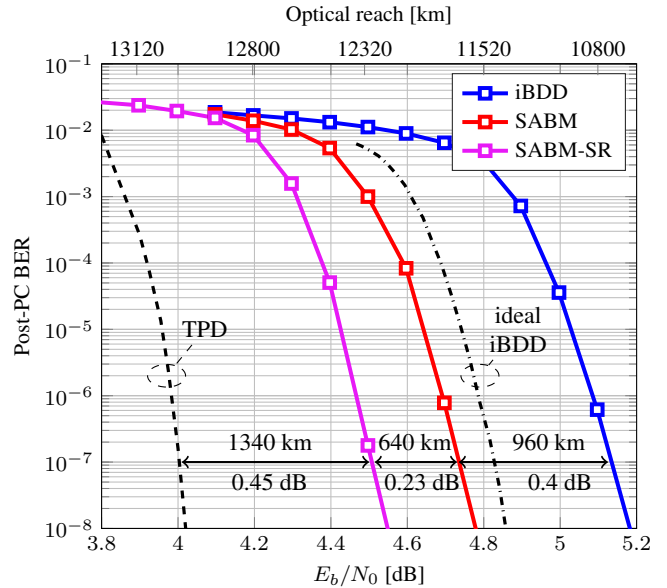

Fig.~\ref{fig:Results2} shows the performance of a PC using a BCH (8,2,1) with rate $R=0.87$. It can be observed that the gains are slightly reduced compared to Fig.~\ref{fig:Results1}, due to the increased code rate. SABM (red line) gain over iBDD (blue line) is now reduced to 0.4 dB, equivalent to a 960 km reach increase. SABM-SR gains 0.23 dB over SABM and 0.63 dB compared to iBDD, equivalent to a 640 km and 1600 km reach extension, respectively. Despite being a \emph{full} SD decoder, TPD shows \emph{only} a 0.45 dB improvement over SABM-SR, which translates to a 1340 km reach increase.    

\section{Conclusions}
We presented a novel soft-aided hard-decision decoder for product codes based on the principle of marking bits using soft information from the channel. Updating the bit marking at each iteration via scaled reliabilities yields up to 0.3 dB gain (or 960 km reach increase) compared to the standard SABM algorithm. SABM-SR outperforms iBDD by up to 0.8 dB (2400 km reach increase) while preserving its core algebraic structure. We believe this new scheme is a strong candidate for 400G and beyond optical transponders.

\vspace{.9ex}
{\small
\noindent\textbf{Acknowledgements:}
This work has received funding from the European Research Council (ERC) under the European Union's Horizon 2020 research and innovation programme (grant agreement No 757791). G. Liga also gratefully acknowledges the NWO Visitors Travel Grant 040.11.659/6291.
}

\section*{References}


\end{document}